\documentstyle[12pt]{article}
\normalbaselines

\font \fiveboldit             = cmbxti10 at 5pt
\font \sevenboldit             = cmbxti10 at 7pt

\font \tenboldit             = cmbxti10 at 10pt
\font \twelveboldit             = cmbxti10 scaled \magstep 1

\newfam\bolditfam
\textfont\bolditfam=\twelveboldit\scriptfont\bolditfam=\sevenboldit
\scriptscriptfont\bolditfam=\fiveboldit
\def\boldit{\fam\bolditfam\twelveboldit}
\def\fboldit{\fam\bolditfam\tenboldit}
\newfont{\fslboldit}{cmbxti10 scaled\magstep2}
\newfont{\slboldit}{cmbxti10 scaled\magstep3}

\font \fivesans               = cmss10 at 5pt
\font \sevensans              = cmss10 at 7pt
\font \tensans                = cmss10

\font\boldit=cmbxti10 scaled \magstep1
\newcommand{\stackscript}[1]{\mathop{\smash{#1}\vphantom{#1}}\limits}

\newcommand{\lstop}[1]{\left.}
\newcommand{\rstop}[1]{\right.}

\def\sqr#1#2{{\vcenter{\hrule height.#2pt
    \hbox{\vrule width.#2pt height#1pt \kern#1pt
      \vrule width .#2pt}
    \hrule height.#2pt}}}

\newcounter{exercise}

\def\theindex{\@restonecoltrue\if@twocolumn\@restonecolfalse\fi
\columnseprule \z@
\columnsep 35pt\twocolumn[\@makeschapterhead{Index}]
 \@mkboth{INDEX}{INDEX}\thispagestyle{plain}\parindent\z@
 \parskip\z@ plus .3pt\relax\let\item\@idxitem
\addcontentsline{toc}{subsection}{Index}}

\def\bbbr{{\rm I\!R}} 

\def\bbbone{{\mathchoice {\rm 1\mskip-4mu l} {\rm 1\mskip-4mu l}
{\rm 1\mskip-4.5mu l} {\rm 1\mskip-5mu l}}}
\def\bbbc{{\mathchoice {\setbox0=\hbox{$\displaystyle\rm C$}\hbox{\hbox
to0pt{\kern0.4\wd0\vrule height0.9\ht0\hss}\box0}}
{\setbox0=\hbox{$\textstyle\rm C$}\hbox{\hbox
to0pt{\kern0.4\wd0\vrule height0.9\ht0\hss}\box0}}
{\setbox0=\hbox{$\scriptstyle\rm C$}\hbox{\hbox
to0pt{\kern0.4\wd0\vrule height0.9\ht0\hss}\box0}}
{\setbox0=\hbox{$\scriptscriptstyle\rm C$}\hbox{\hbox
to0pt{\kern0.4\wd0\vrule height0.9\ht0\hss}\box0}}}}
\def\bbbe{{\mathchoice {\setbox0=\hbox{\smalletextfont e}\hbox{\raise
0.1\ht0\hbox to0pt{\kern0.4\wd0\vrule width0.3pt height0.7\ht0\hss}\box0}}
{\setbox0=\hbox{\smalletextfont e}\hbox{\raise
0.1\ht0\hbox to0pt{\kern0.4\wd0\vrule width0.3pt height0.7\ht0\hss}\box0}}
{\setbox0=\hbox{\smallescriptfont e}\hbox{\raise
0.1\ht0\hbox to0pt{\kern0.5\wd0\vrule width0.2pt height0.7\ht0\hss}\box0}}
{\setbox0=\hbox{\smallescriptscriptfont e}\hbox{\raise
0.1\ht0\hbox to0pt{\kern0.4\wd0\vrule width0.2pt height0.7\ht0\hss}\box0}}}}
\def\bbbq{{\mathchoice {\setbox0=\hbox{$\displaystyle\rm Q$}\hbox{\raise
0.15\ht0\hbox to0pt{\kern0.4\wd0\vrule height0.8\ht0\hss}\box0}}
{\setbox0=\hbox{$\textstyle\rm Q$}\hbox{\raise
0.15\ht0\hbox to0pt{\kern0.4\wd0\vrule height0.8\ht0\hss}\box0}}
{\setbox0=\hbox{$\scriptstyle\rm Q$}\hbox{\raise
0.15\ht0\hbox to0pt{\kern0.4\wd0\vrule height0.7\ht0\hss}\box0}}
{\setbox0=\hbox{$\scriptscriptstyle\rm Q$}\hbox{\raise
0.15\ht0\hbox to0pt{\kern0.4\wd0\vrule height0.7\ht0\hss}\box0}}}}
\def\bbbt{{\mathchoice {\setbox0=\hbox{$\displaystyle\rm
T$}\hbox{\hbox to0pt{\kern0.3\wd0\vrule height0.9\ht0\hss}\box0}}
{\setbox0=\hbox{$\textstyle\rm T$}\hbox{\hbox
to0pt{\kern0.3\wd0\vrule height0.9\ht0\hss}\box0}}
{\setbox0=\hbox{$\scriptstyle\rm T$}\hbox{\hbox
to0pt{\kern0.3\wd0\vrule height0.9\ht0\hss}\box0}}
{\setbox0=\hbox{$\scriptscriptstyle\rm T$}\hbox{\hbox
to0pt{\kern0.3\wd0\vrule height0.9\ht0\hss}\box0}}}}
\def\bbbs{{\mathchoice
{\setbox0=\hbox{$\displaystyle     \rm S$}\hbox{\raise0.5\ht0\hbox
to0pt{\kern0.35\wd0\vrule height0.45\ht0\hss}\hbox
to0pt{\kern0.55\wd0\vrule height0.5\ht0\hss}\box0}}
{\setbox0=\hbox{$\textstyle        \rm S$}\hbox{\raise0.5\ht0\hbox
to0pt{\kern0.35\wd0\vrule height0.45\ht0\hss}\hbox
to0pt{\kern0.55\wd0\vrule height0.5\ht0\hss}\box0}}
{\setbox0=\hbox{$\scriptstyle      \rm S$}\hbox{\raise0.5\ht0\hbox
to0pt{\kern0.35\wd0\vrule height0.45\ht0\hss}\raise0.05\ht0\hbox
to0pt{\kern0.5\wd0\vrule height0.45\ht0\hss}\box0}}
{\setbox0=\hbox{$\scriptscriptstyle\rm S$}\hbox{\raise0.5\ht0\hbox
to0pt{\kern0.4\wd0\vrule height0.45\ht0\hss}\raise0.05\ht0\hbox
to0pt{\kern0.55\wd0\vrule height0.45\ht0\hss}\box0}}}}
\def\bbbz{{\mathchoice {\hbox{$\sans\textstyle Z\kern-0.4em Z$}}
{\hbox{$\sans\textstyle Z\kern-0.4em Z$}}
{\hbox{$\sans\scriptstyle Z\kern-0.3em Z$}}
{\hbox{$\sans\scriptscriptstyle Z\kern-0.2em Z$}}}}
\def\qed{\ifmmode\sq\else{\unskip\nobreak\hfil
\penalty50\hskip1em\null\nobreak\hfil\sq
\parfillskip=0pt\finalhyphendemerits=0\endgraf}\fi}
\newfam\sansfam
\textfont\sansfam=\tensans\scriptfont\sansfam=\sevensans
\scriptscriptfont\sansfam=\fivesans
\def\sans{\fam\sansfam\tensans}

\newtheorem{thm}{Theorem}[section]
\newtheorem{corr}[thm]{Corollary}
\newtheorem{lemma}[thm]{Lemma}
\newtheorem{defin}[thm]{Definition}

\newcommand{\hilb}[1]{\mbox{$\cal #1$}}

\newcommand{\path}[1]{\mbox{$\cal #1$}}

\newcommand{\Operator}[1]{\mathchoice
   {\mbox{\boldmath $#1$}}{\mbox{\boldmath $#1$}}
   {\mbox{\footnotesize \boldmath $#1$}}
   {\mbox{\footnotesize \boldmath $#1$}}} 

\newcommand{\notion}[1]{\em #1\/\em}
\newcommand{\Notion}[1]{{\boldit #1}}
\newcommand{\stress}[1]{{\bf #1}}

\newcommand{\extra}[2]{
 \begin{minipage}{#1} \vspace*{5mm}
 #2 \vspace*{0.5cm}
 \end{minipage} }

\def\Note#1{\medskip \noindent\stress{#1}}
\def\fNotion#1{{\fboldit #1}}

\def\sqr#1#2{{\vcenter{\hrule height.#2pt
    \hbox{\vrule width.#2pt height#1pt \kern#1pt
      \vrule width .#2pt}
    \hrule height.#2pt}}}

\def\Note#1{\medskip \noindent\stress{#1}}

\def\Op#1{{\Operator #1}}

\begin{document}

\title{\hfill {\normalsize ASI-TPA/021/95}\linebreak\vskip 2mm
Axiomatic Quantum Theory\thanks{Extended version of a lecture
delivered at the \notion{Second German-Polish Symposium}, Zakopane,
Poland, Part I: School of Physics, 5--10 September 1995.}}

\date{October 1995}
\author{W.\ L\"ucke\thanks{e-mail: aswl@pta1.pt.tu-clausthal.de}\\
Arnold Sommerfeld Institute for Mathematical Physics\\
Technical University of Clausthal\\
Leibniz Str.\ 10, D-38678 Clausthal, Federal Republic of Germany}

\maketitle

\def\logicS{{\cal L}}
\def\logic{{(\logicS,\prec,\neg)}}
\def\csublogic{{({\cal L}_{\rm c},\prec,\neg})}
\def\csublogicS{{{\cal L}_{\rm c}}}
\def\clogic{{({\cal B},\prec_B,\neg_B})}
\def\states{{\cal S}}
\def\densop{{\Op T}}
\def\trace{{\rm trace\,}}
\def\lattice{{(\logicS,\prec)}}
\def\SF{{\Operator\Phi}}
\def\vNalg{{\cal M}}
\def\vNlogic{\left(\logicS_\vNalg,\prec,\neg\right)}
\def\lvNalg{\vNalg({\cal O})}
\def\lalg{{\cal A}({\cal O})}
\def\dcones{{\cal K}}
\def\rego{{{\cal O}}}
\def\supp{{\rm supp\/}}
\def\rpg{{\cal P}_+^\uparrow}
\def\qlalg{{\cal A}}

\begin{quote}
``It is difficult and perhaps still somewhat controversial to summarize
the tenets of quantum physics.'' \cite{HaagFI}
\end{quote}

\section{Introduction} \label{SIntro}

The aim of the following lecture is to give some rough overview over
most essential structures underlying all working quantum theoretical
models as well as axiomatic and algebraic quantum field theory.

Before specializing to ordinary Hilbert space quantum theory it will be
explained why common sense reasoning cannot be applied naively and the
pragmatic procedure (\notion{quantum reasoning}) of getting along with
this situation is briefly described. Then the more special structure
underlying ordinary quantum theory will be postulated rather than
derived, since
\begin{quote}
`It is not yet possible to deduce the present form of quantum mechanics
from completely plausible and natural axioms.'' \cite[p.\
62]{MackeyMFQM}
\end{quote}
Within this framework equivalence between the quantum logical and the
algebraic formulation will be established. Finally, the Kastler-Kastler
theory of local observables and Wightman's axiomatic field theory will
be indicated.

Unfortunately, due to lack of time, essential issues like canonical
quantization, GNS represention, or spontaneous symmetry breaking cannot
be discussed.
\medskip

\section{Basic Logical Structure} \label{SBA}

\subsection{Fundamental Postulates} \label{SFP}

Every known concrete quantum theory of closed systems is a
\notion{statistical theory} of the following type. It is affiliated with
\begin{enumerate}
\item
A set ${\cal Q}$ of {\bf
macroscopic} prescriptions for preparing a `state' of the system under
consideration.
\item
A set ${\cal X}$ of {\bf macroscopic} prescriptions for performing
\notion{simple} `tests' (called {\em questions\/} by Piron) on the
system under consideration,\footnote{We do not require that these tests
can be performed  within an arbitrarily small time interval.} i.e.\
tests with only two possible outcomes referred to as `yes' or `no'.
\item

A mapping\footnote{Actually -- as well known for open systems
\cite{Davies} --  the probability for the outcome `yes' or `no' in a
test  performed before the `state' state is prepared need not have any
meaning. However for all known models of {\bf closed} quantum systems
the `states' can be imagined as having been prepared as early as one
likes. This is essential for standard scattering theory.}
$$
{\rm pr\,}:\, {\cal Q}\times{\cal X} \longrightarrow [0,1]
$$
with the following interpretation:\footnote{We do {\bf not} claim that $S$
uniquely characterizes a {\bf microscopic} state, nor do we claim that
$T$ fixes the microscopic details of a test!}
\begin{quote}
${\rm pr\,}(S,T)$ is the probability for the outcome `yes' when performing a
simple `test' corresponding to $T$ on the system in a `state'
corresponding to $S\,$.
\end{quote}
\end{enumerate}
Obviously, the `tests' $T\in{\cal X}$ cannot separate elements
$S_1,S_2\in{\cal Q}\,$, which are equivalent in the following sense:
$$
S_1\sim S_2 \stackrel{\rm def}{\Longleftrightarrow} {\rm
pr\,}(S_1,T)={\rm pr\,}(S_2,T)\;\forall\,T\in{\cal X}\,.
$$
Similarly the `states' $S\in{\cal Q}$ cannot separate `tests' $T_1,T_2\in{\cal
X}\,$, which are equivalent in the sense that
$$
T_1\sim T_2 \stackrel{\rm def}{\Longleftrightarrow} {\rm
pr\,}(S,T_1)={\rm pr\,}(S,T_2)\;\forall\,S\in{\cal Q}\,.
$$
Therefore the appropriate mathematical formalism deals with the
equivalence classes $[S]$ (also called {\em states\/}) and  $[T]$ (also
called {\em propositions\/} or {\em questions}) together with the
(consistent) assignment
$$
\omega\left(\Op P\right)\stackrel{\rm def}{=} {\rm pr\,}(S,T)
\quad \mbox{for } \omega=[S]\,,\; \Op P=[T]
$$
rather than the specific prescriptions $S,T$ and the mapping pr\,.
\smallskip

\noindent
${\cal Q}$ and ${\cal X}$ are always (more or less implicitly) chosen
such that the following three conditions are fulfilled:\footnote{These
conditions are designed to allow for classical reasoning as far as possible.
Implicit in (I$_3$) and (I$_1$) is the following {\em
standardization postulate\/}:
\begin{quote}
For every $\Op P\in\logicS\setminus\left\{\Op 0\right\}$ there exist a
state $\omega\in\states$ with $\omega(\Op P)=1$ and a state
$\omega'\in\states$ with $\omega'(\Op P)=0\,$.
\end{quote}
Therefore semi-transparent windows, e.g., cannot be used for simple tests.}
\begin{itemize}
\item[(I$_1$):]
For every $\Op P \in \logicS\stackrel{\rm def}{=} \left\{ [T]:\,
T\in {\cal X}\right\}$ there is also an element $\neg \Op P \in
\logicS$ fulfilling\footnote{A more general framework was suggested in
\cite{MielnikNLQ}.}
$$
\omega(\neg \Op P) = 1 - \omega(\Op P) \hspace{3mm} \mbox{
for all } \omega \in \path S\stackrel{\rm def}{=} \left\{ [S]:\, S\in
{\cal Q}\right\}\,.
$$
\item[(I$_2$):]
Let  $\Op P_1,\Op P_2, \ldots \in \logicS $. Then there is
an element $\Op I \in \logicS$ such that for all $\omega \in \path S$
$$
\omega( \Op I) = 1 \mbox{ if and only if } \omega( \Op P_j) = 1
\mbox{ for } j=1,2, \ldots
$$
\item[(I$_3$):]
Let  $\Op P_1,\Op P_2, \ldots \in \logicS $ be such that
$$
\omega ( \Op P_j ) = 1 \Longrightarrow \omega ( \Op P_k ) =
0  \hspace{3mm} \mbox{ for all } \omega \in \path S \,\hspace{3mm}
\mbox{ whenever } j < k\,.
$$
Then there is an element $\Op S \in \logicS$ fulfilling
$$
\omega(\Op S)=\omega ( \Op P_1 )+\omega ( \Op P_2 )+\ldots \quad
\forall\,\omega\in\states\,.
$$
\end{itemize}

Note that $(I_1)$ defines a mapping $\neg:\, \logicS\longrightarrow
\logicS\,$. Moreover, there is always a natural semi-ordering of the
elements of $\logicS$ given by
\begin{equation} \label{semi}
P_1 \prec P_2 \stackrel{\rm def}{\Longleftrightarrow}
\omega(P_1)\leq\omega(P_2)\;\forall\, \omega \in{\cal S}\,.
\end{equation}
\vskip 5mm

\begin{thm}[Structure Theorem] \label{TST}
If  $\logicS \neq \emptyset$ and $\path S$ fulfill conditions
(I$_1$)---(I$_3$), then\/ $\left(\logicS ,\prec, \neg\right)\,$, with
$\prec$ given by (\ref{semi}) and $\neg$ given by (I$_1$), is a
\Notion{logic}, i.e.\ a $\sigma$-complete weakly modular lattice $\left({\cal
L} ,\prec\right)\,$. Moreover, under these
conditions, every $\omega \in \path S \;$ is a probability measure
over\/ $(\logicS ,\prec,\neg)$ fulfilling the \Notion{Jauch-Piron condition}
\begin{equation} \label{JauchPiron}
\Bigl(\,\omega\left(\Op P_1 \right)=1=\omega\left(\Op P_2 \right)
\;\Longrightarrow\; \omega\left(\Op P_1\land\Op P_2\right)=1
\,\Bigr)\quad\forall\,\Op P_1\,,\,\Op P_2\in\logicS \,.
\end{equation}
\end{thm}
\vskip 5mm

{\bf Proof:} See \cite[appendix]{DoLuQ} (see also \cite{Macz74} for
related results). \rule[-2mm]{2mm}{4mm}
\vskip 5mm

\noindent
Motivated by the structure theorem, we rely on the following
\vskip 5mm

\begin{center}
\fbox{\parbox{13cm}{{\bf Axiom 1.} Every physical system can be modeled
by some logic $\logic$ in the following way:
\begin{itemize}
\item[(i)]
For every preparable statistical state there is a probability measure
$\omega$ on $\logic$ fulfilling (\ref{JauchPiron}) and for every
performable simple test there is an element $\Op P\in\logicS$ such that
$$
\omega(\Op P) = \,\mbox{probability for the result `yes'}\,.
$$
\item[(ii)]
For all $\Op P_1\,,\,\Op P_2\in\logicS$ we have
$$
\Op P_1=\Op P_2 \;\Longleftrightarrow \; \left( \omega(\Op P_1)=\omega(\Op
P_2)\;\forall\,\omega\in\states\right)\,,
$$
where $\states$ denotes the set of all probability measures
$\omega$ on $\logic$ corresponding to preparable statistical states.
\item[(iii)]
Given $\omega\in\states$ and $\Op P\in\logicS$ with
$\omega(\Op P)\ne0\,$, there is a unique $\omega_{,\Op P}\in\states$
fulfilling\footnotemark
$$
\omega_{,\Op P}(\Op P') = \omega(\Op P')/\omega(\Op P)\quad\forall\,\Op
P'\prec\Op P\,.
$$
\end{itemize}
}}
\end{center}
\footnotetext{Conditions (iii) and (ii) imply that for every $\Op
P\in\logicS\setminus\left\{\Op 0\right\}$ there is a {\em state\/}
$\omega\in\states$ with $\omega(\Op P)=1$ -- which would also be a
consequence of (I$_3$) and (I$_1$).
Naively interpreted, $\omega_{,\Op P}(\Op P')$ describes the
\notion{conditional probability} in the state $\omega$ for $E_{\Op P'}$
-- defined by (\ref{properties}) --
being true provided $E_{\Op P}$ is true. In ordinary quantum theory
$\densop_{\omega_{,\Op P}}$ is given by $\Op P\densop_\omega\Op
P/\omega(\Op P)\,$.}
\vskip 5mm

\noindent
\vskip 5mm

\begin{corr} \label{CAx1}
If Axiom 1 is fulfilled, the following statements hold:
\begin{enumerate}
\item
For all $\Op P\in\logicS$ and $\omega\in\states:$
$$
\omega(\Op P)=1 \Longrightarrow \omega_{,\Op P}=\omega\,.
$$
\item
For every atom\footnote{See Appendix \ref{SCL} for the definition of an
\Notion{atom} $\Op Z\,$. Obviously, $\omega_{\Op Z}$ is a {\em pure
state\/}.} $\Op Z$ of $\logic$ there is a unique $\omega_{\Op
Z}\in\states$ with $\omega_{\Op Z}(\Op Z)=1\,$.
\item
Defining
$$
\omega_{,\Op P}(\Op P')/\omega(\Op P) \stackrel{\rm def}{=} 0 \quad\mbox{\bf
if } \omega(\Op P)=0\,,
$$
we have for all {\bf compatible} $\Op P\,,\,\Op P'\in\logic$
\begin{equation} \label{clstat}
\omega(\Op P') = \omega(\Op P)\,\omega_{,\Op P}(\Op P') + \omega(\neg\Op
P)\, \omega_{,\neg\Op P}(\Op P')\quad\forall\,\omega\in\states
\end{equation}
and
\begin{equation} \label{condprob2}
\omega_{,\Op P_1,\Op P_2} = \omega_{,\Op P_1\land\Op
P_2}\quad\forall\,\omega\in\states\,.
\end{equation}
\end{enumerate}
\end{corr}
\vskip 5mm

\noindent
Just for simplicity we add the following assumption, fulfilled in
ordinary quantum theory:\footnote{In orthodox quantum theory
(\ref{states}) is a consequence of Footnote 3 and $\sigma$-convexity.}
\begin{equation} \label{states}
\fbox{$\states =\,\mbox{set of {\bf all} probability measures on }\logic\,.$}
\end{equation}
\medskip

\subsection{Terminology} \label{SST}

A semi-ordered set $\lattice$ ({\em poset\/}) is called a
\Notion{lattice\/}, if\footnote{For quantum logic: $E_{\Op P_1\land
\Op P_2}$ is certain if and only if both $E_{\Op P_1}$ {\bf and} $E_{\Op
P_2}$ are certain.} both
$$
\Op P_1\land\Op P_2 \stackrel{\rm def}{=} {\textstyle \inf_{{\cal
L}}}\left\{\Op P_1,\Op P_2\right\} \quad \mbox{and} \quad \Op P_1\lor\Op
P_2 \stackrel{\rm def}{=} {\textstyle \sup_{\logicS}}\left\{\Op P_1,\Op
P_2\right\}
$$
exist for arbitrary $\Op P_1,\Op P_2\in\logicS\,$. If
$$
\Op 0 \stackrel{\rm def}{=} \inf \logicS
$$
exists, we say $\lattice$ has a \Notion{universal lower bound\/}
$\Op 0\,$. If
$$
\Op 1 \stackrel{\rm def}{=} \sup \logicS
$$
exists, we say $\lattice$ has a \Notion{universal upper bound\/}
$\Op 1\,$. A bijection $\neg:\, \logicS \longrightarrow \logicS$ is
called an \Notion{orthocomplementation\/} of $\lattice$ if the
latter has universal upper and lower bounds and the
following requirements are fulfilled for arbitrary $\Op P,\Op
P'\in\logicS:$
\begin{quote}
\begin{itemize}
\item[(O$_1$):]
$\Op P\land\neg\Op P=\Op 0\,,$
\item[(O$_2$):]
$\Op P\lor\neg\Op P=\Op 1\,,$
\item[(O$_3$):]
$\neg\left(\neg \Op P\right) = \Op P\,,$
\item[(O$_4$):]
$\Op P\prec\Op P'\Longrightarrow\neg \Op P'\prec\neg\Op P\,.$
\end{itemize}
\end{quote}
An orthocomplemented lattice $\logic$ is called \Notion{distributive\/},
if\footnote{Note that, thanks to orthocomplementation, also
$$
\Op P_1 \lor \left(\Op P_2\land\Op P_3\right) = \left(\Op P_1 \lor \Op
P_2\right)\land  \left(\Op P_1 \lor \Op P_3\right)
$$
holds for all $\Op P_1,\Op P_2,\Op P_3\in\logicS\,$, if (\ref{distrib})
does.}
\begin{equation} \label{distrib}
\Op P_1 \land \left(\Op P_2\lor\Op P_3\right) = \left(\Op P_1 \land \Op
P_2\right)\lor  \left(\Op P_1 \land \Op P_3\right)
\end{equation}
holds for all $\Op P_1,\Op P_2,\Op P_3\in\logicS\,$.

\noindent
An {\bf ordered} pair$(\Op P_1,\Op P_2)\in\logicS\times\logicS$ is
called \Notion{compatible\/} if\footnote{In a logic $\logic$ the ordered
pair $(\Op P_1,\Op P_2)$ is compatible if and only the sublogic
generated by $\Op P_1$ and $\Op P_2$ is classical (see \cite[Definition
(2.14) and Theorem (2.19)]{Piron}). Also this shows that $(\Op P_1,\Op P_2)$ is
compatible if and only $(\Op P_2,\Op P_1)$ is compatible.}
\begin{equation} \label{compat}
\Op P_1 = \left(\Op P_1\land\Op P_2\right)\lor
\left(\Op P_1\land\neg\Op P_2\right)\,.
\end{equation}
An {\bf ordered} pair$(\Op P_1,\Op P_2)\in\logicS\times\logicS$ is
called \Notion{modular} if (\ref{distrib}) holds for all $\Op
P_3\prec\Op P_1\,$, i.e.\ if
\begin{equation} \label{modpair}
\logicS\ni\Op P_3 \prec \Op P_1 \Longrightarrow \Op P_1 \land \left(\Op
P_2\lor\Op P_3\right) = \left(\Op P_1 \land \Op P_2\right)\lor \Op
P_3\,.
\end{equation}
An orthocomplemented lattice $\logic$ is called \Notion{weakly
modular\/}\,\footnote{It is called \Notion{modular\/}, if all pairs
$\left(\Op P_1,\Op P_2\right)\in \logicS\times \logicS$ are modular.} if
$$
\Op P_1\perp\Op P_2 \Longrightarrow \left(\Op P_1,\Op P_2\right) \mbox{
modular}\,,
$$
holds for all $\Op P_1,\Op P_2\in\logicS\,$, where
$$
\Op P_1\perp\Op P_2 \stackrel{\rm def}{\Longleftrightarrow} \left(\Op
P_1\prec \neg\Op P_2\right)\,.
$$
It is easy to prove, that $({\cal
L},\prec,\neg)$ is weakly modular if and only if
$$
\Op P_2\prec \Op P_3 \Longrightarrow \left(\Op P_2\,,\,\Op P_3\right) \mbox{
compatible}
$$
holds for all $\Op P_2\,,\,\Op P_3\in \logicS\,$.

$({\cal L},\prec,\neg)$ is called an \Notion{orthomodular lattice\/} if
it is a weakly modular orthocomplemented lattice.
\vskip 5mm

{\footnotesize
\begin{quote}
{\bf Remark:} An orthocomplemented lattice $({\cal
L},\prec,\neg)$ is weakly modular if and only if
$$
\left(\Op P_2\,,\,\Op P_3\right) \mbox{ compatible } \Longleftrightarrow
\left(\Op P_3\,,\,\Op P_2\right) \mbox{ compatible}
$$
holds for all $\Op P_2\,,\,\Op P_3\in \logicS\,$. $({\cal
L},\prec,\neg)$ \cite[Theorem 21, p.\ 53]{Birkhoff}.
\end{quote}}
\vskip 5mm

A lattice $\lattice$ is called ($\sigma$-)\Notion{complete} \footnote{As
shown by Mac Neille, every poset can be embedded into a complete lattice
in such a way that semi-ordering is preserved as well as greatest lower
bounds and lowest upper bounds existing in the poset \cite[Theorem
V.22]{Birkhoff}.} if both $\inf_\logicS M$ and $\inf_\logicS M$ exist
for every (countable) $M\subset\logicS\,$.

A \Notion{logic} is a $\sigma$-complete orthomodular
lattice.\footnote{This is the definition is equivalent to that given in
\cite[p.\ 105]{Vara1}. There are also others adding conditions on the
set of states (see, e.g., \cite{Pulma}).}

A logic $\logic$ is called \Notion{classical} (or \notion{Boolean
algebra}) if it is distributive.\footnote{We do not require classical
logics to be atomic (see Appendix \ref{SCL}).} Otherwise it is called a
\Notion{quantum logic}.

A \Notion{probability measure\/} on the logic $\logic$ is a mapping
$\omega:\,\logicS \longrightarrow[0,1]$ fulfilling the following two
conditions:\footnote{Varadarajan calls such $\omega$ just {\em measures\/}
on $\logic$ \cite[p.\ 113]{Vara1}, unless $\logic$ is classical.}
\begin{quote}
\begin{itemize}
\item[(S$_1$):]
$\omega(\Op 1) = 1\,,$
\item[(S$_1$):]
For every countable subset $\left\{\Op P_1\,,\,\Op P_2\,,\ldots\right\}
\subset \logicS$ we have
$$
\omega\left({\textstyle \sup_\logicS}\left\{ \Op P_j:\,
j=1,2,\ldots\right\}\right) = \sum_{j=1}^\infty\omega\left(\Op P_j\right)
\mbox{\bf \ if } \Op P_j\perp\Op P_k \mbox{ for } j\ne k\,.
$$
\end{itemize}
\end{quote}
\medskip

\subsection{Some Warnings} \label{SSW}

In `orthodox quantum mechanics' \cite{Primas} (without superselection
rules\footnote{A system modeled by $\logic$ is said to possess
\Notion{superselection rules} if the \Notion{center}
$$
{\cal C}\logic \stackrel{\rm def}{=}\left\{\Op P\in\logicS:\, (\Op P,\Op
P') \mbox{ compatible } \forall\,\Op P'\in\logicS\right\}
$$
of $\logic$ is nontrivial $(\logicS\ne{\cal C}\ne\left\{ \Op 0,\Op
1\right\}$). $\logic$ is called \Notion{irreducible} if ${\cal
C}=\left\{ \Op 0,\Op 1\right\}\,$.}) the logic $\logic$ described in
Section \ref{SFP} is realized as follows ({\em standard quantum
logic\/}):
\begin{itemize}
\item
$\logicS$ is given as the set of all projection operators\footnote{Their
specific physical identification depends on the \notion{dynamics}, as
discussed in \cite{MielnikNLQ} and \cite{LueckeNL}.} in some separable
complex Hilbert space $\hilb H\,$ of dimension $\geq 2\,$.
\item
For arbitrary $\Op P_1,\Op P_2\in \logicS$ we have
$$
\begin{array}[c]{rcl}
\Op P_1\prec\Op P_2 &\stackrel{\rm def}{\Longleftrightarrow}& \Op
P_1\leq\Op P_2\\
&\Longleftrightarrow& \left( \left\langle \Psi \mid
\Op P_1 \Psi \right\rangle \leq \left\langle \Psi \mid
\Op P_1 \Psi \right\rangle\;\forall\,\Psi\in\hilb H\right)\,.
\end{array}
$$
\item
For every $\Op P\in\hilb H$ we have
$$
\neg \Op P \stackrel{\rm def}{=} \Op 1 -\Op P \,.
$$
\end{itemize}
If ${\rm dim\,}(\hilb H)\geq 3\,$, by Gleason's theorem \cite{Gleason},
for every $\omega\in\states$ there is a unique positive trace class
operators\footnote{Conversely, every trace class operator of trace 1
induces a probability measure on standard quantum logic.}
$\densop_\omega\in\hilb B(\hilb H)$ fulfilling
$$
\omega(\Op P) = \trace\left(\densop_\omega \Op P\right)
\quad\forall\,\Op P\in\logicS\,.
$$
{}From this it is easily seen that {\bf none} of the $\omega\in\states$
can be \Notion{dispersion free\/}\,, i.e.\ fulfill the requirement
$$
\omega(\Op P)\in\left\{0,1\right\}\quad\forall\,\Op P\in\logicS
$$
(not even approximately). For a very long time this was taken as
evidence for nonexistence of {\em hidden variables\/} -- even though
D.\ Bohm constructed a consistent (nonlocal) hidden variable theory in the
beginning of the fifties \cite{BohmHV}.
Actually, in order to avoid this conclusion one has to abandon the
seemingly natural assumption that {\bf every microscopic state} --
not only those given by $\states$ -- induces a probability
measure on $\logic\,$.

\noindent
However, also from a hidden variables point of view, this assumption has
to be questioned:
\vskip 5mm

\vbox{
\begin{lemma}[D.\ Pfeil] \label{LPfeil}
For every set $\hat\logicS$ there is a {\bf classical}
logic $({\cal B}, \prec_B,\neg_B)$ and a
mapping $M:\,\hat\logicS\longrightarrow{\cal B}$ for
which the following holds:
\begin{quote}
For every mapping $\omega:\, \hat\logicS \longrightarrow [0,1]$ there is
a probability measure $\mu$ on $({\cal B}, \prec_B,\neg_B)$ fulfilling
$$
\omega(\Op P)=\mu\Bigl(M(\Op P)\Bigr)\quad\forall\,\Op P\in\hat\logicS\,.
$$
\end{quote}
\end{lemma}}
\vskip 5mm

{\bf Proof:} In order to avoid purely technical
complications\footnote{The general proof is by straightforward adaption
of a construction given by Kochen and Specker \cite[Section
I]{KochSpeck}.} we
consider only the case of finite
$$
\hat\logicS = \left\{ \Op P_1,\ldots,\Op P_n \right\}\,.
$$
Then we may take
$$
\begin{array}[c]{c}
B\stackrel{\rm def}{=}\left\{0,1\right\}^{n}\;,\quad \prec_B\stackrel{\rm
def}{=} \subset\,,\\
\neg_BM'\stackrel{\rm def}{=}B\setminus M'\quad\mbox{for } M'\subset
B\,,\\
M(\Op P_\nu)\stackrel{\rm def}{=}\left\{ b=(b_1,\ldots,b_n)\in B:
\,b_\nu=1 \right\}\quad\mbox{for } \nu\in\left\{1,\ldots,n\right\}\,,
\end{array}
$$
and
$$
\mu_\omega(M')\stackrel{\rm def}{=}\sum_{(b_1,\ldots,b_n)\in
M'}\prod_{\nu=1}^{n} \omega_\nu(b_\nu)\quad\mbox{for }
\omega\in\states\,,\,M'\subset B\,,
$$
where
$$
\omega_\nu(1)\stackrel{\rm def}{=}1-\omega_\nu(0)\stackrel{\rm def}{=}
\omega(\Op P_\nu)\quad\mbox{for } \nu\in\left\{1,\ldots,n\right\}\,.
\qquad\rule[-2mm]{2mm}{4mm}
$$
\vskip 5mm

\noindent
It seems natural to assign `actual' properties $E_{\Op P}$ to the elements
of $\logicS$ in the sense that:\footnote{Assumption (iii) of Axiom 1
then says that for every $\Op P\in\logicS$ there is a state
$\omega\in\states$ in which $E_{\Op P}$ is certain.}
\begin{equation} \label{properties}
\begin{array}[c]{l}
\mbox{A system in the state $\omega\in {\cal S}$ has property $E_{\Op P}$ {\bf
with
certainty} if and}\\
\mbox{only if $\omega(P)=1\,$.}
\end{array}
\end{equation}
We are used giving names to these properties like `spin up', `positive
energy' and so on. However, from the proof of Lemma \ref{LPfeil} it
should be clear that there is no evidence\footnote{From a `hidden
variables' point of view the `test' enforces a transition, if necessary,
of the system to a state in which either $E_{\Op P}$ or $E_{\neg\Op P}$ is
certain. Typically, for micro-systems, the number of cases in which the
criteria for this alternative are not specified by $T$ cannot be
neglected, causing apparent indeterminism with respect to incompatible
properties $E_{\Op P_1},E_{\Op P_2}\,$. According to the Kopenhagen
interpretation of quantum mechanics, indeterminism is a direct
consequence of the hypotheses, never accepted by Einstein, that quantum
theory presents a {\bf complete} description of physical reality.} for
the assumption that under all circumstances -- independent of any test -- the
system has
either property $E_{\Op P}$ or property $E_{\neg \Op P}$ -- even
though\footnote{We doubt that more detailed specification of the
measurement context might be of any help, here.}
$$
\omega(\neg \Op P) = 1 -\omega(\Op
P)\quad\forall\,\omega\in\states\,,\,\Op P\in \logicS
$$
and even though tests corresponding to $\Op P$ and $\neg\Op P$ can
typically be performed jointly. Therefore, it is no surprise that we
encounter quantum peculiarities such as\footnote{The set of `states'
determines the (quantum logical) relations between (equivalence classes
of) tests, which must not be interpreted too naively.}
\begin{equation} \label{qpec1}
\omega(\Op P)=1 \not\Longrightarrow \left( \omega(\Op P\land\Op P')=
\omega(\Op P') \; \forall\,\Op P'\in\logicS\right)
\end{equation}
or\footnote{By (\ref{JauchPiron}), $\Op P_1\land \Op P_2=0$ means that
there is no \underline{preparable} property guaranteeing both $E_{\Op P_1}$
and $E_{\Op P_2}\,$.}
\begin{equation} \label{qpec2}
\Op P_1\land\Op P_2=0 \not\Longrightarrow \Op P_1 \prec \neg \Op P_2\,.
\end{equation}
\vskip 5mm

\noindent
 Let us call a
state $\omega\in\states$ \Notion{classical}, if (\ref{clstat}) holds for
all pairs $\Op P\,,\,\Op P'\in\logic$ -- whether compatible or not.
\vskip 5mm

\vbox{\begin{lemma} \label{Lclassic}
A $\sigma$-complete orthocomplemented lattice $\logic$ fulfilling
conditions (i)--(iii) of Axiom 1 for (\ref{states}) is a classical logic, if
and only if all $\omega\in\states$ are classical.
\end{lemma}}
\vskip 5mm

{\bf Proof:}  See Appendix \ref{SCS}. \rule[-2mm]{2mm}{4mm}

\medskip

\subsection{Quantum Reasoning} \label{QR}

In spite of all warnings, \Notion{simple quantum reasoning\/} according
to the following rules is consistent:
\begin{itemize}
\item
Choose a {\bf classical} sublogic $\csublogic$ of $\logic$ and forget
about all the other elements of $\logicS\,$.
\item
Then imagine that every {\bf individual} -- in whatever situation -- has
either property $E_{\Op P}$ or $E_{\neg\Op P}$ {\bf if} ${\Op P}\in
\csublogicS\,$.
\item
For $\omega\in\states\,$, imagine that $\omega(\Op P)$ is the relative
number of individuals having property $E_{\Op P}$ in an ensemble
corresponding to $\omega$ {\bf if} ${\Op P}\in\csublogicS\,$.
\item
Imagine that $\prec$ corresponds to common sense logical implication and
that $\neg$ corresponds to common sense logical negation.
\end{itemize}
\vskip 5mm

\noindent
This way all quantum peculiarities are avoided. For instance, in spite
of (\ref{qpec1}), we may
conclude
$$
\lstop\{\begin{array}{l}
\omega(P_1) = 1\,,\\
P_1 \mbox{ compatible with } P_2
\end{array}\right\}
\;\Longrightarrow\; \omega(P_1\land P_2) = \omega (P_2)\;\forall\,\omega
$$
or even
$$
\begin{array}[c]{l}
P_1\,,\; P_2 \mbox{ compatible\footnotemark}\\
\Longrightarrow\; \omega(P_1\lor P_2) = \omega(P_1\land \neg P_2) +
\omega(\neg P_1\land P_2) + \omega(P_1\land P_2)\;\forall\,\omega\,.
\end{array}
$$
\footnotetext{$$
\begin{array}[c]{rcl}
P_1\lor P_2 = (P_1\lor P_2)\land 1 &=& (P_1\land \neg P_2)\lor P_2\\
&=& (P_1\land \neg P_2)\lor(\neg P_1\land P_2)\lor(P_1\land P_2)
\end{array}
$$}

\noindent
Simple quantum reasoning naturally leads to the notion of {\em
observable\/}:\footnote{The Borel ring on $\bbbr^1$ could be replaced by
an arbitrary classical logic.}
\vskip 5mm

\begin{defin} \label{DObserv} 
An \Notion{observable} $A$ of a physical system modeled by the logic
$\logic$  is a $\sigma$-morphism $\Op E_A$ of the Borel ring on the real line
into $\logic$ which is unitary, i.e.\ $\Op E_A(\bbbr)=\Op 1\,$.
\end{defin}
\vskip 5mm

\noindent
The physical interpretation of $\Op E_A$ in the sense of quantum
reasoning is as follows:
\begin{quote}
Given $\omega\in\states$ and a Borel subset $\Delta$ of $\bbbr^1$ then
$\omega\bigl(\Op E_A(\Delta)\bigr)$ can be imagined as the relative
number of individuals for which $A\in\Delta$ in an ensemble corresponding
to $\omega\,$.
\end{quote}
Consequently, the expectation value\footnote{Of course, the expectation
value may be infinite!} for $A$ in an ensemble
corresponding to $\omega$ is given by the Stieltjes integral
\begin{equation} \label{expect}
\overline{A}(\omega) = \int \lambda \,{\rm d}\omega\left(\Op
E_A\left((-\infty,\lambda]\right)\right) \,.
\end{equation}
In orthodox quantum theory $\Op E_A$ is a projection valued
measure\footnote{Of course, in general, tests corresponding to the $\Op
E_A(\Delta)$ can never be exactly realized. Therefore many people prefer
to use just positive operator valued measures.}
and (\ref{expect}) can also be written as
\begin{equation} \label{expect'}
\overline{A}(\omega) = \trace\left( \densop_\omega \Op A \right)
\end{equation}
where
\begin{equation} \label{spectres}
\Op A \stackrel{\rm def}{=} \int \lambda \,{\rm d}\Op
E_A\left((-\infty,\lambda]\right)
\end{equation}
is a self-adjoint operator ({\em spectral representation\/}).
\vskip 5mm

\noindent
Simple quantum reasoning can be applied to a whole family observables
$A_1,A_2,\ldots$ if and only if all the pairs
$$
\left(\Op E_{A_j}(\Delta_j),\Op E_{A_k}(\Delta_k)\right)\;, \quad
\Delta_j\,,\,\Delta_k\in\bbbr
$$
are compatible. For \Notion{bounded} $A_j\,$, i.e.\ if $\Op
E_{A_j}(\Delta_j)=1$ for suitable compact $\Delta_j\in\bbbr\,$, this is
equivalent to pairwise commutativity of the corresponding (bounded)
self-adjoint operators $\Op A_j\,$.
\vskip 5mm

\noindent
In order to make predictions for multiple tests one has to know how states
change as a result of a simple test. Here we assume\footnote{Usually, a
test causes a much more drastic change of the state or even ends by
absorbing the corresponding individual. An ideal test, typically, would
be approximately realized by a highly efficient filter.}
$$
\fbox{\parbox{13cm}{\Notion{L\"uders' Postulate:} For every $\Op
P\in\logicS$ there is a corresponding \notion{ideal test} causing a
transition\footnotemark\ $\omega\mapsto\omega_{,\Op P}$ whenever the result
is `yes'.}}
$$
\footnotetext{Remember the second statement of Corollary \ref{CAx1},
however. If $\Op Z_1$ and $\Op Z_2$ are atoms of $\logic\,$,
$\omega_{\Op Z_1}(\Op Z_2)$ is called the {\em transition probability\/}
for the transition $\omega_{\Op Z_1} \longrightarrow \omega_{\Op
Z_2}\,$.}
\begin{quote}
{\footnotesize{\bf Remark:} The L\"uders postulate ensures that an {\bf
ideal} test corresponding to $\Op P$ destroys none of the properties
$E_{\Op P'}$ with $\Op P'\,,\,\Op P$ compatible.

{\bf Proof:} Let $\Op P'\,,\,\Op P$ be compatible and
$$
\omega(\Op P')=1\;,\quad \omega(\Op P)\ne 0\,.
$$
Then
$$
1 = \omega(\Op P') \stackscript{=}_{(\ref{clstat})} \omega(\Op P)
\omega_{\Op P}(\Op P') + \left(1-\omega(\Op P)\right)\omega_{\neg\Op P}(\Op P')
$$
implies $\omega_{\Op P}(\Op P')=1\,$. \rule[-1.5mm]{1.5mm}{3mm}
}
\end{quote}
\vskip 5mm

\noindent
By L\"uders' postulate,\footnote{In the relativistic theory L\"uders'
postulate causes interesting problems \cite{Schlieder} (see also
\cite{MittelRQL},\cite{MitSta}).} given the initial state
$\omega\in\states\,$, the probability for the \notion{homogeneous
history} $(\Op P_1,\ldots,\Op P_n)$ -- i.e.\ for getting the answer
`yes' for all subsequent ideal tests of a series corresponding to $\Op
P_1,\ldots,\Op P_n\in\logicS$ -- should be\footnote{Naively interpreted,
$\omega(\Op P_1)\omega_{,\Op P_1}(\Op P_2)\cdots\omega_{,\Op P_1,
\ldots,\Op P_{n-1}}(\Op P_n)$ is the probability for joint validity of
the properties $E_{\Op P_1},\ldots,E_{\Op P_1}$ in the state $\omega\,$.
Usually (see, e.g., \cite{OmnesIQM}, \cite{GriffithQR}), unfortunately, this is
formulated in the Schr\"odinger picture, thus imposing unnecessary
restrictions.}
$$
\omega(\Op P_1)\omega_{,\Op P_1}(\Op P_2)\cdots\omega_{,\Op P_1,
\ldots,\Op P_{n-1}}(\Op P_n)\,.
$$
Consistent quantum reasoning with respect to histories leads to the
modern notion of {\em decoherent histories\/}.
\vskip 5mm

\noindent
Given a history $(\Op P_1,\ldots,\Op P_n)$ not corresponding to a simple
test, we can no longer be sure that there is an initial state for which
$(\Op P_1,\ldots,\Op P_n)$ is certain, i.e., for which $\omega(\Op
P_1)\omega_{,\Op P_1}(\Op P_2)\cdots\omega_{,\Op P_1, \ldots,\Op
P_{n-1}}(\Op P_n)=1\,$. Therefore the `logic' of histories is weaker
than that for simple tests and may provide a useful basis for
generalizing quantum theory \cite{IshamQLDH}.
\medskip

\subsection{Dynamics} \label{SDyn}

\vbox{\begin{defin} \label{DSymm}
A \Notion{symmetry} of a physical system modeled\footnotemark by the
logic $\logic$ is an automorphism of $\logic\,$, i.e.\ a bijection of
$\logicS$ onto itself preserving the least upper bound and the
orthocomplementation.
\end{defin}}
\footnotetext{If (\ref{states}) does not hold one should also require
$\alpha^*(\states)=\states$  and then the inverse of a symmetry need not
be a symmetry (A.\ Bohm's point of view).}
\vskip 5mm

\noindent
In orthodox quantum theory, by Wigner's theorem \cite[\S3--2]{Piron},
for every automorphism $\alpha$ of $\logic$ there is an operator $\Op
V\in\hilb B(\hilb H)$ which is either unitary or anti-unitary and
fulfills
\begin{equation} \label{Wigsymm}
\alpha(\Op P) = \Op V \Op P\Op V^*\quad \forall \Op
P\in\logicS\subset\hilb B(\hilb H)\,.
\end{equation}

\noindent
For simplicity we consider only those systems 
for which time translation is a
symmetry:\footnote{
Note that the dual of a symmetry has always an inverse in the set of all
probability measures. In this sense evolution can always be extrapolated
backwards in time!}
$$
\fbox{\parbox{13cm}{{\bf Axiom 2.} For every time $t$ there is a
symmetry $\alpha_t$ with the following physical interpretation:
\begin{quote}
Let $T$ be a macroscopic prescription for performing a simple test
corresponding to $\Op P\in\logicS\,$. Then the prescription $T_t$ to do
everything prescribed by $T$ just with time delay $t$ characterizes a
test corresponding to $\alpha_t(\Op P)\,$.
\end{quote}
$t\longmapsto \alpha_t$ is \Notion{weakly continuous}, i.e., for fixed
$\Op P\in\logicS$ and $\omega\in\states$ the probability
$\omega\left(\alpha_t(\Op P)\right)$ is a {\bf continuous} function of
$t\,$.}}
$$
\vskip 5mm

\noindent
The family $\left\{ \alpha_t \right\}_{t\in\bbbr}$ determines the
\Notion{dynamics} of the system. According to its definition it has to
be a weakly continuous 1-parameter group of transformations:
\begin{equation} \label{1pgt}
\alpha_0={\rm id}\,,\qquad \alpha_{t_1}\circ\alpha_{t_2} =
\alpha_{t_1+t_2} \quad\forall\,t_1\,,\,t_2\in\bbbr\,.
\end{equation}
For orthodox quantum mechanics this implies that there is a self-adjoint
operator, the \Notion{Hamiltonian}, $\Op H$ on $\hilb H$
fulfilling
\begin{equation} \label{undyn}
\alpha_t(\Op P) = e^{it\Op H}\Op Pe^{-it\Op
H}\quad\forall\,t\in\bbbr\,,\,\Op P\in\logicS,,
\end{equation}
but:
\begin{quote}
``...we omit its surprisingly
difficult proof, which involves some theorems about group cocycles.''
\cite[p.\ 26]{Davies}
\end{quote}
(see Section \ref{SSymm} and Appendix \ref{SSWL} for a sketchy proof).

\noindent
The Hamiltonian is unique only up to an additive multiple of the
identity operator. But, in any case, it has to be bounded from below.
This has been exploited in a nice, easy to prove, theorem by Hegerfeldt
which, unfortunately, caused a lot of irritation\footnote{See, e.g.,
B.\ Schroer: \notion{Reminiscences about Many Pitfalls and Some
Successes of QFT Within the Last Three Decades}, hep-th/9410085, pp.\
7--8; to appear in Reviews in Mathematical Physics.} in connection with some
misleading application to Fermi's two-atoms problem \cite{Hegerprl}:
\vskip 5mm

\vbox{\begin{thm}[Hegerfeldt] \label{THegerfeld}
Let $\Op H$ be a self-adjoint operator on $\hilb H$ which is semibounded
from below. Moreover let $\Op P$ be a nonnegative bounded operator on
$\hilb H$ and $\Psi\in\hilb H\,$. Then either
$$
\left\langle \Psi \mid e^{i\Op Ht}\Op Pe^{-i\Op Ht}\Psi \right\rangle =
0 \quad\forall\,t\in\bbbr
$$
or
$$
t_1\ne t_2 \Longrightarrow
\int_{t_1}^{t_2} \left\langle \Psi \mid e^{i\Op Ht}\Op Pe^{-i\Op Ht}\Psi
\right\rangle > 0\,{\rm d}t\quad\forall\,t_1\,\,t_2\in\bbbr\,.
$$
\end{thm}}
\vskip 5mm

\Note{Remark:} Even if $\logic$ is isomorphic to the standard quantum
logic one may use a nonlinear realization of $\logic$ in Hilbert space
such that time evolution has to be described by a group of nonlinear
transformations \cite{LueckeNL} -- in full agreement with Mielnik's
program for handling nonlinear Schr\"odinger equations \cite{MielnikNLQ}.
\medskip

\section{General Quantum Theory} \label{SOQT}

Usually one is only concerned with suitable sublogics of standard
quantum logic. Therefore we assume the following:\footnote{The
$C^*$-algebraic approach may be considered as a preliminary step: One has
to find a suitable representation (superselection structure) and take
the weak closure in this representation to get the von Neumann algebra.}
\vskip 5mm

\begin{center}
\fbox{\parbox{13cm}{{\bf Axiom 3.} There is a {\bf
separable}\footnotemark\ Hilbert space $\hilb H$ and a {\bf von Neumann
algebra} $\vNalg\subset\hilb B(\hilb H)$ by which $\logic$ is realized
in the following way:
\begin{itemize}
\item
$\logicS=\logicS_{\vNalg}\stackrel{\rm def}{=}\left\{\Op P\in\vNalg:\,
\Op P^*=\Op P=\Op P^2\right\}\,$,
\item
$\Op P_1\prec\Op P_2 \stackrel{\rm def}{\Longleftrightarrow} \Op
P_1\leq\Op P_2\quad\forall \, \Op P_1\,,\,\Op P_2\in\logicS\,$,
\item
$\neg\Op P \stackrel{\rm def}{=}\Op 1 -\Op P\quad\forall\,\Op P\in\logicS\,$.
\end{itemize}}}
\end{center}
\footnotetext{For an interesting application of nonseparable Hilbert space
see, e.g., \cite{BuchQED}.}
\medskip

\subsection{Von Neumann Algebras} \label{SVNA}

A \Notion{von Neumann algebra} is a
subalgebra $\vNalg$ of $\hilb B(\hilb H)\,$, $\hilb H$ some Hilbert
space, that is given by the \Notion{commutant}
$$
{\cal N}' \stackrel{\rm def}{=}\left\{ \Op A\in\hilb B(\hilb H):\, [\Op
A, \Op B]_-=0\;\forall\,\Op B\in {\cal N}\right\}
$$
of some $*$-invariant subset ${\cal N}\subset\hilb B(\hilb H):$
$$
\vNalg = \left({\cal N}\cup{\cal N}\right)'\qquad \left(\mbox{and hence
}\vNalg=\vNalg''\right)\,.
$$
The projection operators of a von Neumann algebra always form a sublogic
$\vNlogic$ of the corresponding standard quantum logic.\footnote{Here,
$\Op{P}_{1} \wedge \Op{P}_{2} =  \; \mbox{s-} \lim_{n
\rightarrow \infty} (\Op{P}_{1}\Op{P}_{2})^{n}\,$. Necessary
and sufficient conditions for a quantum logic to be isomorphic to a
sublogic am standard quantum logic are given in \cite{GudderEQL}.} This
does not hold for arbitrary $C^*$-algebras.\footnote{For example, the
$C^{*}$-subalgebra of $\hilb{L}(\hilb{H})$ generated by $\Op{1}$
and $\hilb{C}(\hilb{H})$ contains exactly those projection operators
$\Op{P}$ for which either $\Op{P}$ itself or $\Op{1} -
\Op{P}$ has finite rank. This cannot be consistent with
$\sigma$-completeness, required for a logic.}
\vskip 5mm

\noindent

A von Neumann algebra $\vNalg$ is called a \Notion{factor} if its
\Notion{center}
$$
{\cal Z}(\vNalg) \stackrel{\rm def}{=} \vNalg\cap\vNalg''
$$
is trivial, i.e., if ${\cal Z}(\vNalg)=\bbbc\Op 1\,$. This equivalent to
$\vNlogic$ being irreducible,\footnote{Recall Footnote 17.} i.e.\
to the absence of \Notion{superselection rules}.
\vskip 5mm

\noindent
A factor of \Notion{type I$_n$}, $n\in\left\{2,\ldots,\infty\right\}\,$,
is the von Neumann algebra $\hilb B(\hilb H)$ on a Hilbert space $\hilb H$ of
dimension $n\,$, i.e., the corresponding logic is a standard quantum logic.
\medskip

\subsection{State Functionals} \label{SSF}

A \Notion{state functional} on a von Neumann algebra $\vNalg$ is a mapping
$\Op{A} \rightarrow \omega (\Op{A})$ of $\vNalg$ into the
\stress{complex} numbers fulfilling the following three conditions:
$$
\begin{array}{rll}
(S_{1}): & \omega(\Op{A} + \alpha \Op{B}) =
\omega(\Op{A}) + \alpha \omega(\Op{B}) &
(\mbox{\Notion{linearity}}) \\
(S_{2}): & \omega(\Op{1}) = 1 & (\mbox{\Notion{normalization}}) \\
(S_{3}): & \omega(\Op{A}^{*}\Op{A}) \geq 0 &
(\mbox{\Notion{positivity}\,\footnotemark})
\end{array}
$$
\footnotetext{It would be quite tedious to show in general for
$C^{*}$-algebras that  $\Op{A}^{*}\Op{A} = -
\Op{B}^{*}\Op{B} \Longrightarrow
\Op{A}^{*}\Op{A} = 0 \,$.}

\noindent
One can easily show that the following three conditions conditions are
fulfilled for every state $\omega$ on $\vNalg:$
$$
\begin{array}{rll}
(i) & | \omega(\Op{A}^{*}\Op{B}) | ^{2} \leq
\omega(\Op{A}^{*}\Op{A})
\omega(\Op{B}^{*}\Op{B}) & (\mbox{\Notion{Cauchy Schwarz
inequality}}) \\
(ii) & \omega(\Op{A}^{*}) = \overline{\omega (\Op{A})} &
(\mbox{\Notion{hermiticity}}) \\
(iii) & \omega(\Op{A}_{n}) \rightarrow \omega (\Op{A}) \; \;
\mbox{ if } \| \Op{A} - \Op{A}_{n} \| \rightarrow 0 &
(\mbox{\Notion{continuity}})
\end{array}
$$
\begin{lemma} \label{L2.1.5}
Let $\vNalg \subset \hilb{B}(\hilb{H})$ be a von Neumann algebra
and let $\left\{ \Op{A}_{i} \right\}_{i \in I} \subset \vNalg$ be
an increasing net of positive operators fulfilling $\sup_{i \in
I} \| \Op{A}_{i} \| < \infty\,$. Then $\sup_{i \in
I}\Op{A}_{i}$ exists with respect to $\hilb{B}(\hilb{H})$ and is an
element of the algebra $\vNalg\,$.
\end{lemma}
\medskip

{\bf Proof:} See \cite[Lemma 2.4.19]{BraRob1}. \rule [-2mm]{2mm}{4mm}
\vskip 5mm

\noindent
A state functional of the von Neumann algebra $\omega$ on $\vNalg$ is called
\Notion{normal},\footnote{In
general, a state $\omega$ on a von Neumann algebra is called
{\fboldit singular}, if for every nonzero projection operator
$\Op{P}$ there is another nonzero projection operator
$\Op{P'}$ for which $\omega (\Op{P'}) = 0$ and
$\Op{P'} \prec \Op{P}\,$. An example for such a state is
$$
\omega(\Op A) \stackrel{\rm def}{=} \lim_{N \to \infty}
\sum_{\nu =1}^N \trace\left( \Op P_\nu \Op A \right) \, ,
$$
where the $\Op P_\nu$ denote rank-1 projection operators
corresponding to a complete orthonormal basis of $\hilb H\,$.} if
$$
\omega (\sup_{i \in I} \Op{A}_{i}) = \sup_{i \in I} \omega(
\Op{A}_{i})
$$
holds for every net fulfilling the requirements of Lemma \ref{L2.1.5}.
The following two theorems shows that the normal state functionals are
the algebraic equivalent to the probability measures.
\vskip 5mm

\vbox{\begin{thm}[generalized Gleason theorem] \label{TGGT}
Let $\vNalg$ be a von Neumann algebra with no type I$_2$
summand.\footnotemark\
Every finitely additive probability measure $\omega$ on
$\logicS_\vNalg$ can be extended to a \Notion{state functional} on
$\logicS_\vNalg\,$. This state functional is normal if and only if
the corresponding probability measure is completely additive.
\end{thm}}
\footnotetext{Of course, the theorem cannot be true if $\vNalg$
is a type I$_2$ factor since then
$$
\omega(\Op 1-\Op P)=1-\omega(\Op P)\geq 0\quad\forall\,\Op P\in
\logicS_\vNalg
$$
is the only requirement for a probability measure $\omega\,$.}
\vskip 5mm

{\bf Proof:} See \cite{Maeda}. \rule[-2mm]{2mm}{4mm}
\vskip 5mm

\vbox{\begin{thm} \label{TNS}
A state $\omega$ on the von Neumann algebra $\vNalg$ is normal if and only if
there is a positive trace class operator\footnotemark\ $\Op T_\omega\in\hilb
B(\hilb H)\supset\vNalg$ with
$$
\omega(\Op A) = \trace\left(\Op T_\omega\Op A\right)\quad\forall\,\Op
A\in\vNalg\,.
$$
\end{thm}}
\footnotetext{Note, however, that $\Op T_\omega$ is no longer unique, in
general.}
\vskip 5mm

{\bf Proof:} \cite[Theorem 2.4.21]{BraRob1}. \rule[-2mm]{2mm}{4mm}
\vskip 5mm

\noindent
{}From now on we identify the states with their corresponding normal state
functionals. Then the bounded\footnote{The self-adjoint operators
corresponding to unbounded observables of $\vNlogic$ are said to be
{\fboldit affiliated} to $\vNalg\,$. An operator $\Op A$ on $\hilb H$ is
affiliated to $\vNalg$ if and only if $\Op U\Op A\Op U'= \Op A$ holds
for all unitary operators $\Op U\in\vNalg'$ \cite[I, 1, exerc.\
10]{Dix1}.} observables $A$ can be identified with the self-adjoint
elements $\Op A\in\vNalg$ in the sense of (\ref{expect'}).
\medskip

\subsection{Symmetry Groups} \label{SSymm}

Thanks to the generalized Gleason theorem one can show the following:
\vskip 5mm

\begin{corr} \label{C1PSG}
Let $\vNalg$ be a von Neumann algebra and $\left\{ \alpha_t
\right\}_{t\in\bbbr}$ a weakly continuous 1-para\-me\-ter group of
symmetries of $\vNlogic\,$. Then $\left\{ \alpha_t \right\}_{t\in\bbbr}$
is the restriction to $\logicS_\vNalg$ of a weakly$^*$
continuous\footnote{$\left\{ \alpha_t \right\}_{t\in\bbbr}$ is
\Notion{weakly$^*$ continuous} iff $\omega\left(\alpha_t(\Op A)\right)$
is continuous in $t$ for all normal states $\omega$ and all $\Op
A\in\vNalg$ \cite[Propositon 2.4.3]{BraRob1}.} 1-parameter group of
$C^*$-automorphisms\footnote{A $C^*$-automorphism $\alpha$ of $\vNalg$
is a linear automorphism $\alpha$ of $\vNalg$ fulfilling the conditions
$$
\alpha(\Op{A}\Op{B}) = \alpha(\Op{A})
\alpha(\Op{B}) \quad \forall\,\Op{A}\,,\,\Op{B} \in
\vNalg\,,
$$
and
$$
\alpha(\Op{A}^{*}) = \left( \alpha(\Op{A})\right)^*\quad
\forall\,\Op{A}\in\vNalg\,.
$$} of $\vNalg\,$.
\end{corr}
\vskip 5mm

{\bf Proof:}  See Appendix \ref{SSWL}. \rule[-2mm]{2mm}{4mm}
\vskip 5mm

\noindent
Therefore, the weakly continuous 1-parameter groups of
\Notion{symmetries} of a system modeled by $\vNlogic$ can be identified
with the weakly$^*$ continuous 1-parameter groups of $C^*$-automorphisms of
$\vNalg\,$. The most important 1-parameter group of symmetries is the
time translation symmetries. If $\vNalg$ is of type I these
$C^*$-automorphisms are generated in the standard way by some (in
general unbounded) Hamiltonian $\Op H:$
\vskip 5mm

\begin{thm} \label{T2.1.15}
Let $\{ \alpha_{t} \}_{t \in \bbbr^{1}}$ be a weakly$0*$ continuous one
parameter group of
$C^{*}$-automorphisms of \hilb{L}(\hilb{H})\/. Then there is a unique
self-adjoint operator $\Op{H}$ fulfilling
$$
\alpha_{t}(\Op{A}) = e^{\frac{i}{\hbar}\Op{H}t}
\Op{A} e^{-\frac{i}{\hbar}\Op{H}t} \hspace{5mm}
\forall \Op{A} \in \hilb{L}(\hilb{H})\, , \; t \in \bbbr^{1}\,.
$$
\end{thm}
\medskip

{\bf Proof:} See \cite[Example 3.2.35]{BraRob1} and Stone's
theorem. \rule [-2mm]{2mm}{4mm}
\medskip

\section{Relativistic Quantum Theory} \label{SRQT}

For a relativistic quantum theory Axiom 2 has to be enhanced:
$$
\fbox{\parbox{15cm}{{\bf Axiom 2'.} There is a weakly$^*$ continuous
representation of $\rpg$ by $C^*$-automorphism
$\alpha_{\Lambda,x}$ of $\vNalg\,$, with obvious interpretation
generalizing that of the dynamics
$$
\alpha_t \stackrel{\rm def}{=}\alpha_{\bbbone,(t,0,0,0)}\quad\mbox{ for
} t\in\bbbr\,.
$$}}
$$

\subsection{Algebras of Local Observables} \label{SALO}

Let $\lvNalg$ denote the subalgebra of $\vNalg$ generated by
all those $\Op P\in\logicS_\vNalg$ corresponding to tests that can be
performed within the space-time region
$$
\rego\in\dcones\stackrel{\rm def}{=}\left\{\mbox{open double
cones}\right\}\,.
$$
This identification implies\footnote{Usually {\bf all} projection
operators in $\lvNalg$ are considered as corresponding to tests
performable within $\rego\,$. However, this is not as evident as tacitly
assumed in standard presentations like \cite{HaagLQP} since it implies
that the interpretation of $\Op A$ and $\omega$ depends on the selected
space-time region $\rego:$ The unit operator considered as an element of
$\lvNalg$ has to be identified with the maximal equivalence class $\Op
1_\rego$ of simple tests performable within $\rego$ and
$\omega\in\states$ has to be {\bf locally} interpreted via
$$
\omega(\Op A) = \frac{{\rm counting \; rate\; for\;}\Op A}{{\rm
counting \; rate\; for\;}\Op 1_\rego}\,.
$$}
$$
\rego_1\subset\rego_1\Longrightarrow\vNalg(\rego_1)\subset\vNalg(\rego_2)
\qquad\forall\,\rego_1\,,\,\rego_2\in\dcones
$$
and
$$
\alpha_{\Lambda,x}\left(\vNalg(\rego)\right)=\vNalg\left(
\Lambda\rego+x\right)\quad\forall\,(\Lambda,x)\in\rpg\,,\,\rego\in\dcones\,.
$$

\noindent
According to \cite{HaKa} one also assumes
$$
\vNalg = \displaystyle\left(\bigcup_{\rego\in{\cal
K}}\lvNalg\right)''\,.
$$
Then the following Problem arises:
\begin{quote}
$\vNalg$ depends on the representation of
$\displaystyle\bigcup_{\rego\in{\cal K}}\lvNalg\,$, but only tests
corresponding to $\Op P\in\lvNalg$ are realistic, i.e., only
$[\lvNalg]\,,\;\rego\in{\cal K}\,,$ can be experimentally determined (in
principle).
\end{quote}
\vskip 5mm

\noindent
The solution of this problem is based on the following
\underline{Haag-Kastler assumptions:}
\smallskip

\noindent
There is a separable Hilbert space $\hilb H\,$, a continuous unitary
representation\footnote{Physical interpretation:
$$
\Op U(A,x)\pi_0(\Op B)\Op U(A,x)^{-1}= \pi_0\left(\alpha_{\Lambda_A,x}(\Op
A)\right) \mbox{ for } \Op B\in\lvNalg\,.
$$} $\Op U(A,x)$ of iSL(2,$\bbbc)$ in $\hilb H\,$, and a faithful
irreducible representation $\pi_0$ of the
$$
C^*\mbox{-completion of }\bigcup_{\rego\in\dcones}\lvNalg
$$
in $\hilb H$ such that the net of \Notion{local algebras}
$$
\lalg \stackrel{\rm def}{=}\pi_0\left(\lvNalg\right)\,,\;\rego\in\dcones\,,
$$
fulfills the following requirements:
$$
\begin{array}[c]{lcl}
\mbox{\Notion{isotony}}:&&\\
\hfill \rego_1\subset\rego_2&\Longrightarrow&{\cal
A}\left(\rego_1\right) \subset{\cal A}\left(\rego_2\right)\,,\\
\mbox{\Notion{\underline{locality}}}:&&\\
\hfill\rego_1\,\mbox{spacelike}\,\rego_2\;&\Longrightarrow&\;\left[{\cal
A}\left(\rego_1\right) ,{\cal A}\left(\rego_2\right)\right]_-=0\,,\\
\mbox{\Notion{covariance}}:&&\\
\mbox{\phantom{xxxxxx}}\hfill\Op U(A,x)\lalg\Op U(A,x)^{-1} &=&
{\cal A}\left(\Lambda_A(\rego)+x\right)\,,\\
\mbox{\Notion{spectrum condition}}:\footnotemark&&\\
\hfill\supp \widetilde\varphi \cap \overline{V_+}=0 &\Longrightarrow&
\displaystyle\int \Op U(\bbbone,x)\varphi(x)\,{\rm d}x = 0\,,\\
\mbox{\Notion{cyclic vacuum}}:&&\\
\hfill\overline{\bigcup_{\rego\in\dcones}\lalg\,\Omega}=\hilb H\;,&&\Op
U(A,x)\Omega=\Omega\,.
\end{array}
$$
\footnotetext{\fNotion{Fourier transform:}
$\widetilde\varphi(p)\stackrel{\rm def}{=}(2\pi)^{-2}\displaystyle
\int\varphi(x)e^{ipx}\,{\rm d}x\,$.}
\vskip 5mm

Now $\vNalg$ can be constructed following the
\underline{Haag-Doplicher-Roberts approach:}\footnote{See \cite{HaagLQP}
and references given there.}
\begin{enumerate}
\item
Determine the physically relevant irreducible representations $\pi_\mu$
(\Notion{superselection sectors}) of
$$
\qlalg\stackrel{\rm def}{=}
C^*\mbox{-completion of }\bigcup_{\rego\in\dcones}\lalg
$$
via \notion{localized endomorphisms}\footnote{An endomorphism $\rho$ of
$\qlalg$ is called \fNotion{localized} (in $\rego$) if$$
\rho(\Op A)=\Op A\quad\forall\,\Op A\in\rego'\stackrel{\rm def}{=}
\left\{x\in\bbbr^4:\, x \,\mbox{spacelike}\,\rego\right\}\,.
$$} of $\qlalg\,$.
\item
Take
$$
\lvNalg = \bigoplus_{\mu}\pi_\mu\left(\lalg\right)\,.
$$
\end{enumerate}
\vskip 5mm

\noindent
The guiding principle of this approach is:
\begin{quote}
All physical information -- especially the $\Op S$-matrix \cite{Lue83}--
is already encoded in the local net structure!
\end{quote}
\vskip 8mm

\noindent
Now the question arises:
\begin{quote}
How to construct a concrete local net $\left\{\lalg\right\}_{\rego\in\dcones}$
fulfilling the Haag-Kastler assumptions?
\end{quote}
\vskip 5mm

\noindent
In the simplest case $\lalg$ is generated by all
$$
\exp\left(i\int \Op A(x)\varphi(x)\,{\rm
d}x\right)\;,\quad\varphi=\varphi^*\in{\cal S}(\rego)\,,
$$
where $\Op A(x)$ a given (observable) scalar Wightman field.
\medskip

\subsection{Quantum Fields} \label{SSQ}

The observable $\Op A(x)$ (in the distributional sense) of a scalar field
is usually characterized by the following \underline{Wightman axioms}
\cite{StWi}:
$$
\parbox{14cm}{\begin{itemize}
\item[(W0):]
Poincar\'e symmety implemented by continuous unitary representation $\Op
U(\Lambda,x)$ of $\rpg$ on separable Hilbert space $\hilb H\,$.
\item[(W2):]
$\widetilde\varphi \cap \overline{V_+}=0 \Longrightarrow
\displaystyle\int \Op U(\bbbone,x)\varphi(x)\,{\rm d}x = 0\,$.
\item[(W3):]
There is a normed vector $\Omega\in\hilb H\,$, unique up to a phase
factor, fulfilling
$$
\Op U(\bbbone,x)\Omega = \Omega\quad\forall\,x\in\bbbr^4\,.
$$
\item[(W4):]
$\Op A(x)$ is an operator valued generalized function over a space
${\cal T}$ of \notion{test functions} on $\bbbr^4\,$, with invariant dense
domain $D\subset\hilb H\,$;
${\cal T} = {\cal S}(\bbbr^4)\quad$ (Schwartz space of tempered functions).
\item[(W5):] $
\lstop\{\begin{array}[c]{l}
\Op U(\Lambda,a)D =D\,,\\
\Op U(\Lambda,a)\Op A(x)\Op U(\Lambda,a)^{-1} = \Op A(\Lambda x+a)
\end{array}\right\}\quad\forall\,\Op U(\Lambda,a)\,.
$
\item[(W6):]
$x\,\mbox{spacelike}\,y\Longrightarrow[\Op A(x),\Op A(y)]_-=0\,$
\item[(W7):]
$\Omega$ cyclic with respect to the \notion{smeared} field operators
$\Op A(\varphi)=\displaystyle\int\Op A(x)\varphi(x)\,{\rm d}x\,,\,
\varphi\in{\cal T}\,$.
\end{itemize}}
$$
\vskip 5mm

\noindent
For $\Op A(x)$ to be `observable' one has to add the
requirement\footnote{Naive interpretation: $\Op
A(\varphi)$ is the observable for $\int
A(x)\varphi(x)\,{\rm d}x\,$, with $A(x)$ a measurable field.}
$$
\varphi=\varphi^*\;\Longrightarrow\; \Op A(\varphi) \mbox{ essentially
self-adjoint\footnotemark\ on } D\,.
$$
\footnotetext{See \cite{BoZi} for a useful criterion.}
\vskip 5mm

\noindent
Of course, for constructing realistic dynamics, like quantum
electrodynamics, one also needs unobservable fields like Dirac spinors
$\Psi(x)$ or electromagnetic potentials $\Op A^\mu(x)\,$. One even has to
use an indefinite metric if one wants $\Op A^\mu(x)$ to be covariant and
local. The situation becomes even worse in nonlinear gauge theories.
Nevertheless, the Wightman frame with obvious generalization of (W5)
should be adequate for all {\bf observable} fields (with tempered high
energy behaviour). Thus quantum electrodynamics should finally be given
by an observable Wightman tensor field $\Op F^{\mu\nu}(x)$ for the
electromagnetic field strength and a Wightman vector field
$\jmath^\mu(x)$ for the charge-current density. The more regrettable is
the following fact:
\begin{quote}
No Wightman field with nontrivial $\Op S$-matrix is
known\footnote{Even apparently nontrivial models turned out belong to the
Borchers class of generalized free fields \cite{RehrenGFF}.}
(on 4-dimensional space-time)!
\end{quote}
\medskip

\noindent
Maybe the main obstruction comes from the purely technical assumption
${\cal T} = {\cal S}(\bbbr^4)$ implying ad hoc high energy restrictions
\cite{WightHyFu}. Therefore one should use a more general framework
within which the main results of axiomatic field theory are still valid
(\cite{LuHPA},\cite{LuPCT}). As a first step one could try to construct
theories of Efimov's type (see, e.g., \cite{Efimov1}).
\medskip

\appendix

\section{Symmetries of von Neumann logics} \label{SSWL}

As a simple consequence of the generalized Gleason theorem we have the
following
\vskip 5mm

\begin{corr} \label{CAutom}
Let $\vNalg$ be a $W^*$ algebra on the \stress{separable} complex
Hilbert space $\hilb{H}$ and let $\alpha$ be an automorphism of the
corresponding logic $\vNlogic\,$.
Then there is a bijection $\varphi$ of
$\vNalg_{\rm s} \stackrel{\rm def}{=} \left\{ \Op{A}
= \Op{A}^{*} \in \vNalg \right\}$ onto itself
and a bijection $\varphi_{*}$ of
$$
N_\vNalg \stackrel{\rm def}{=} \left\{ \mbox{\stress{normal} states on }
\vNalg \right\}
$$
onto itself fulfilling the following two conditions:
$$
\alpha (\Op{P}) = \varphi (\Op{P}) \quad\forall\,\Op
P\in\logicS_\vNalg\,,
$$
$$
\left( \varphi_{*}\omega \right) \left( \varphi(\Op{A})
\right) = \omega(\Op{A}) \forall\,\omega \in N_\vNalg\,,\,
\Op{A}\in\vNalg_{\rm s}\,.
$$
\end{corr}
\vskip 5mm

\begin{thm} \label{T2.1.12}
Let $\vNalg$ be a von Neumann algebra, let $\varphi$ be a
bijection of $\vNalg_{s} \stackrel{\rm def}{=}
\left\{\Op{A} = \Op{A}^{*} \in \vNalg \right\}$
onto itself, and let $\varphi_{*}$ be a bijection of
\[
N_{\cal M} \stackrel{\rm def}{=} \left\{ \mbox{normal states
on } \vNalg \right\}
\]
onto itself fulfilling
\[
\left( \varphi_{*} \omega \right) \left( \varphi (\Op{A})
\right) = \omega (\Op{A}) \hspace{5mm} \mbox{ for all }
\omega \in N_{\cal M} \mbox{ and } \Op{A} =
\Op{A}^{*} \in \vNalg\, .
\]
Then $\varphi$ has a unique continuation to a \Notion{Jordan automorphism}
of $\vNalg\,$, i.e.\ to a linear bijection $\varphi$ of $\vNalg$ onto
itself fulfilling\footnote{$\Op{A} \circ \Op{B} \stackrel{\rm
def}{=} \frac{1}{2} \{ \Op{A},\Op{B} \}$ is the
so-called \notion{Jordan} \notion{product}.}
\[
\varphi(\Op{A}^{*}) = \left( \varphi(\Op{A})
\right)^{*} \; , \; \; \varphi \left( \{\Op{A},\Op{B}
\} \right) = \left\{ \varphi (\Op{A}), \varphi
(\Op{B}) \right\} \hspace{5mm} \mbox{ for all }
\Op{A},\Op{B} \in \vNalg \, .
\]
\end{thm}
\medskip

{\bf Sketch of Proof:} $\varphi_{*}$ is easily seen
to be \notion{affine}, i.e.\ to fulfill the condition
\[
\varphi_{*} \left( \lambda \omega_{1} + (1 - \lambda) \omega_{2}
\right) = \lambda (\varphi_{*} \omega_{1}) + (1 - \lambda)
(\varphi_{*} \omega_{2})
\]
for all $\lambda \in [0,1]$ and all $\omega_{1}, \omega_{2} \in
N_{\cal M}\,$:
\[
\begin{array}[t]{rcl}
\left[ \varphi_{*} \left(\lambda  \omega_{1} + (1 - \lambda)
\omega_{2} \right) \right] (\varphi \varphi^{-1} \Op{A}) &
= & \left(\lambda  \omega_{1} + (1 - \lambda) \omega_{2} \right)
(\varphi^{-1} \Op{A}) \\
& = & \lambda  \omega_{1} (\varphi^{-1} \Op{A}) + (1 -
\lambda) \omega_{2} (\varphi^{-1} \Op{A}) \\
& = & \lambda (\varphi_{*} \omega_{1}) (\varphi \varphi^{-1}
\Op{A}) + (1 - \lambda) (\varphi_{*} \omega_{2}) (\varphi
\varphi^{-1} \Op{A}) \\
& = & [ \lambda (\varphi_{*} \omega_{1}) + (1 - \lambda)
(\varphi_{*} \omega_{2})](\Op{A}) \hspace{5mm} \forall
\Op{A} \in \vNalg_{s} \, .
\end{array}
\]
According to Bratelli and Robinson \cite[Theorem 3.2.8]{BraRob1} this
implies the statement of Theorem \ref{T2.1.12}. \rule [-2mm]{2mm}{4mm}
\vskip 5mm

\noindent
One easily proves the following:
\vskip 5mm

\begin{lemma} \label{LJordan}
Let $\vNalg \subset \hilb{L}(\hilb{H})$ be a von Neumann algebra
and let $\varphi$ be a Jordan automorphism of $\vNalg\,$. Then
the following statements hold:
\begin{enumerate}
\item
$0 \neq \Op{A} \in \vNalg \Longrightarrow 0 < \varphi
(\Op{A}^{*} \Op{A})\,$.
\item
The restriction of $\varphi$ to $\logicS_\vNalg$ is an automorphism of
$\vNlogic\,$.
\item
$\omega \in N_{\cal M} \Longrightarrow \varphi_{*} \omega \in N_{\cal M}
\;$, where $(\varphi_{*} \omega) (\Op{A}) \stackrel{\rm def}{=}
\omega \left( \varphi^{-1} (\Op{A}) \right)  \; \forall
\Op{A} \in \vNalg\,$.
\end{enumerate}
\end{lemma}
\vskip 5mm

Now we see that symmetries of a von Neumann logic $\vNlogic$  are in
1-1-corres\-pon\-dence to the Jordan automorphisms of $\vNalg\,$. At
least for factors $\vNalg$ the latter have to be either
$C^{*}$-automorphisms or $C^{*}$-antiautomorphisms:
\vskip 5mm

\begin{thm} \label{T2.1.14}
Let $\varphi$ be a Jordan automorphism of the von Neumann algebra
$\vNalg\,$. Then there is a $\Op P\in\logicS_\vNalg\cap\vNalg'$ for which
$$
\alpha(\Op A)\stackrel{\rm def}{=}\varphi(\Op A)\Op P
$$
is a {\boldmath $C^{*}$}-\Notion{automorphism} of $\vNalg\,$, i.e.\  a Jordan
automorphism fulfilling
$$
\alpha(\Op{A}\Op{B}) = \alpha(\Op{A})
\alpha(\Op{B}) \hspace{5mm} \forall \Op{A},\Op{B} \in
\vNalg\, ,
$$
and
$$
\widehat\alpha(\Op A)\stackrel{\rm def}{=}\varphi(\Op A)(\Op 1 -\Op P)
$$
a {\boldmath $C^{*}$}-\Notion{antiautomorphism} of $\vNalg\,$, i.e.\ a Jordan
automorphism fulfilling
$$
\widehat\alpha(\Op{A}\Op{B}) =
\widehat\alpha(\Op{B}) \widehat\alpha(\Op{A}) \hspace{5mm}
\forall \Op{A},\Op{B} \in \vNalg\, ,
$$
\end{thm}
\medskip

{\bf Proof:} See \cite[Proposition 3.2.2]{BraRob1}.
\rule [-2mm]{2mm}{4mm}
\vskip 5mm

\noindent
Since the Jordan automorphisms for time translation may be
written $\alpha_{t} = \alpha_{t/2} \circ \alpha_{t/2}$ they are
also $C^{*}$-\stress{auto}morphisms,\footnote{For generalizations
see \cite{Roos}.} by Theorem \ref{T2.1.14}.
\vskip 5mm

\noindent
Let us close this section with the following

\extra{13.5cm}{\stress{Warning:} The algebraic formalism
requires the observables to have no physical dimensions,
otherwise addition of observables would make no sense in general.
Hence a system of physical units should be specified to which the
numbers characterized by the observables refer.}
\medskip

\section{Classical States} \label{SCS}

\begin{lemma} \label{LCS}
Let $\logic$ be a $\sigma$-complete orthocomplemented lattice fulfilling
conditions (i)--(iii) of Axiom 1 for (\ref{states}). Then the following
conditions are equivalent:\footnote{To have equivalence of (i) and (ii)
it is sufficient to postulate weak modularity instead of condition (ii) of
Axiom 1 \cite[Theorem (2.15)]{Piron}.}
\begin{itemize}
\item[(i)]
$\logic$ is a classical logic.
\item[(ii)]
$(\Op P_1\land\Op P_2)\lor(\Op P_1\land\neg\Op P_2)\lor(\neg\Op P_1\land\Op
P_2)\lor(\neg\Op P_1\land\neg\Op P_2)=\Op 1\quad \forall\,\Op
P_1\,,\,\Op P_2\in\logicS\,$.
\item[(iii)]
All probability measures $\omega$ on $\logic$ are {\em classical\/}.
\end{itemize}
\end{lemma}
\vskip 5mm

{\bf Proof for $(i)\Longleftarrow(ii):$}\footnote{See also \cite[Theorem
(2.15)]{Piron}.}\\
Since
$$
(\Op P_1\land\Op P_2)\lor(\Op P_1\land\neg\Op P_2)\prec\Op P_1\;,\quad
(\neg\Op P_1\land\Op P_2)\lor(\neg\Op P_1\land\neg\Op P_2)\prec\neg\Op P_1\,,
$$
we have
$$
\omega\left((\Op P_1\land\Op P_2)\lor(\Op
P_1\land\neg\Op P_2)\right)\leq \omega(\Op P_1)\;,\quad
\omega\left((\neg\Op P_1\land\Op P_2)\lor(\neg\Op P_1\land\neg\Op
P_2)\right) \leq \omega(\neg\Op P_1)\,.
$$
for all $\omega\in\states\,$. Since, on the other hand,
$$
\omega\left((\Op P_1\land\Op P_2)\lor(\Op P_1\land\neg\Op
P_2)\right) + \omega\left((\neg\Op P_1\land\Op
P_2)\lor(\neg\Op P_1\land\neg\Op P_2)\right) \stackscript{=}_{(ii)}
1 = \omega(\Op P_1) + \omega(\neg\Op P_1)
$$
for all $\omega\in\states\,$, this implies
$$
\omega\left((\Op P_1\land\Op P_2)\lor(\Op
P_1\land\neg\Op P_2)\right) = \omega(\Op P_1)
$$
for all $\omega\in\states$ and hence
$$
(\Op P_1\land\Op P_2)\lor(\Op P_1\land\neg\Op P_2)= \Op P_1\,,
$$
by condition (ii) of Axiom 1. This means that all pairs $(\Op P_1,\Op
P_2)$ are compatible, hence (i), by Footnote 10. 

{\bf Proof for $(ii)\Longleftarrow(iii):$}\\
Applying (4) to $\omega=\omega_{,\Op P_1}$ we get
$$
\omega_{,\Op P_1}\geq \omega_{,\Op P_1}(\Op P_2)\omega_{,\Op P_1,\Op P_2}\,.
$$
Since, by definition, $\omega_{,\Op P_1}(\neg \Op P_1)=0\,$, therefore
$$
\omega_{,\Op P_1}(\Op P_2)=0\quad\mbox{or}\quad\omega_{,\Op P_1,\Op P_2}(\Op
P_1)=1\,.
$$
Since, by definition, also $\omega_{,\Op P_1,\Op P_2}(\Op P_2)=1\,$, the
{\bf Jauch-Piron condition} (2) gives
\vskip 5mm
$$
\omega_{,\Op P_1}(\Op P_2)=0\quad\mbox{or}\quad\omega_{,\Op P_1,\Op P_2}(\Op
P_1\land\Op P_2)=1\,.
$$
Since $\Op P_1\,,\,\Op P_2$ are arbitrary, we also have:
$$
\begin{array}[c]{rcl}
\left(1 -\omega_{,\Op P_1}(\Op P_2)\right)=0 &\mbox{or}& \omega_{,\Op
P_1,\neg\Op P_2}(\Op P_1\land\neg\Op P_2)=1\,,\\
\omega_{,\neg\Op P_1}(\Op P_2)=0 &\mbox{or}& \omega_{,\neg\Op P_1,\Op P_2}(
\neg\Op P_1\land\Op P_2)=1\,,\\
\left(1 -\omega_{,\neg\Op P_1}(\Op P_2)\right)=0 &\mbox{or}& \omega_{,\neg\Op
P_1, \neg\Op P_2}(\neg\Op P_1\land\neg\Op P_2)=1\,.
\end{array}
$$
By double application of (2):
$$
\begin{array}[c]{rcl}
\omega &=& \omega(\Op P_1)\omega_{,\Op P_1} + \left(1-\omega(\Op
P_1)\right) \omega_{,\neg\Op P_1}\\
&=& \omega(\Op P_1) \left( \omega_{,\Op P_1}(\Op P_2) \omega_{,\Op
P_1,\Op P_2} + \left(1-\omega_{,\Op P_1}(\Op P_2)\right) \omega_{,\Op
P_1,\neg\Op P_2}\right)+\\
&&\left(1- \omega(\Op P_1)\right)\left(\omega_{,\neg\Op P_1}(\Op P_2)
\omega_{,\neg\Op P_1,\Op P_2} + \left(1-\omega_{,\neg\Op P_1}(\Op
P_2)\right) \omega_{,\neg\Op P_1,\neg\Op P_2}\right)\,.
\end{array}
$$
Therefore
$$
\Op P\perp (\Op P_1\land\Op P_2)\lor(\Op P_1\land\neg\Op P_2)\lor
(\neg\Op P_1\land\Op P_2)\lor(\neg\Op P_1\land\neg\Op P_2)
\Longrightarrow \omega(\Op P)=0 \;\forall\,\omega
$$
and hence
$$
\Op P\perp (\Op P_1\land\Op P_2)\lor(\Op P_1\land\neg\Op P_2)\lor
(\neg\Op P_1\land\Op P_2)\lor(\neg\Op P_1\land\neg\Op P_2)
\Longrightarrow \Op P=0\,.
$$
This implies (ii).

{\bf Proof for $(iii)\Longleftarrow(i):$}\\
(i) implies
$$
\omega_{,\Op P}(\Op P') = \omega(\Op P\land\Op P')/\omega(\Op P)\quad
\forall\,\Op P\,,\,\Op P'\in\logicS\,,\,\omega\in\states
$$
and hence (iii). \rule[-2mm]{2mm}{4mm}
\medskip

\section{Covering Law} \label{SCL}

Let $\lattice$ be a lattice with universal lower bound $\Op 0\,$. Given
$\Op P_1\prec\Op P_2\ne\Op P_1\,$, one says that $\Op P_2$
\Notion{covers} $\Op P_1$ if
$$
[\Op P_1,\Op P_2]_\logicS \stackrel{\rm def}{=}\left\{ \Op
P\in\logicS:\, \Op P_1\prec\Op P\prec\Op P_2\right\} = \left\{\Op
P_1,\Op P_2\right\}\,.
$$
$\Op P\in\logicS$ is called an \Notion{atom} of $\lattice$ if it covers
$\Op 0\,$. $\lattice$ is called \Notion{atomic} if every interval $[\Op
0,\Op P]_\logicS$ with $\Op P\ne\Op 0$ contains at least one atom.
\vskip 5mm

\noindent
A logic $\logic$ is said to fulfill the \Notion{covering law} if
\begin{enumerate}
\item
$\lattice$ is atomic.
\item
For all $\Op Z\,,\,\Op P\in\logicS:$
$$
\lstop\{\begin{array}[c]{l}
\Op Z\mbox{ atom}\,,\\
\Op Z\land\Op P=0
\end{array}\right\}\Longrightarrow \Op Z\lor\Op P \mbox{ covers } \Op P\,.
$$
\end{enumerate}
Assuming irreducibility and the covering law in addition to Axiom 1 one
may prove that -- apart from some exceptional cases -- $\logic$ is
isomorphic to the logic of all projection operators on some {\em
generalized Hilbert space\/} \cite[Section 3--1]{Piron}.
\medskip

\section{Quantum Logic via Constraints} \label{SQLvC}

A simple example, given in \cite{DoLuQ}, shows that already the
standardization postulate (Footnote 4) may lead to quantum logic:
\smallskip

Let $M = \{ a_{1},a_{2},b_{1},b_{2} \}$ be a set of 4 elements, $\left(
{\bf B},\subset ,M\setminus. \right)$ the corresponding {\bf classical}
logic of all subsets of $M$, and \path W the set of all probability
measures on $\left( {\bf B},\subset ,M\setminus. \right)\,$.
Restrict now \path W by assuming that only
those $\omega = \omega_{\lambda}$ correspond to
experimentally realizable statistical ensembles, for which there
is a quadruple $\lambda = (\lambda_{1}, \ldots,\lambda_{4})$ of
nonnegative numbers that fulfills the following five
conditions:
\begin{equation} \label{Norm}
\lambda_{1} + \ldots + \lambda_{4} =\frac{1}{2}
\end{equation}
$$
\omega_\lambda( \left\{ a_{1} \right\} ) = \lambda_{1} + \lambda_{4}\,,
\;\omega_\lambda( \left\{ a_{2} \right\} ) = \lambda_{2} +\lambda_{3}\,,
$$
$$
\omega_\lambda( \left\{ b_{1} \right\} ) = \lambda_{1} + \lambda_{2}
\,,\; \omega_\lambda( \left\{ b_{2} \right\} ) = \lambda_{3}+\lambda_{4}\,,
$$
i.e.:
$$
{\cal X} = \{ \omega_\lambda \}\,.
$$
The standardization postulate allows properties $E \in {\cal B}=2^M$ to be
`measurable' only if
\[
\omega_{\lambda}(E) = 1 \mbox{ and }  \omega_{\lambda'}(E) = 0 \mbox{
for suitable } \lambda,\lambda' \mbox{ depending on } E\,.
\]
Then ${\cal Q}$ consists, apart from the empty set $\emptyset$ and the total
set $M\,$, of the following four subsets of $M:$
\[
T_{1}\stackrel{\rm def}{=} \{a_{1},b_{1}\}\,,\;
T_{2}\stackrel{\rm def}{=} \{a_{2},b_{1}\}\,,\;
T_{3}\stackrel{\rm def}{=} \{a_{2},b_{2}\}\,,\;
T_{4}\stackrel{\rm def}{=} \{a_{1},b_{2}\}\,.
\]
Conditions (I$_1$)---(I$_3$) are fulfilled for ${\cal Q}\,,\,{\cal
X}\,$, but $\logic$ is no longer classical.
Indeed, $\logic$ corresponds to the Hasse diagramm
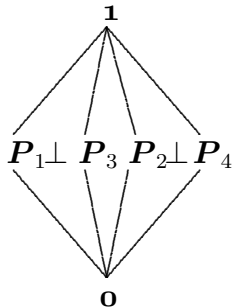
\input prepictex
\input  pictex
\input postpictex
$$
\font\thinlinefont=cmr5
\begingroup\makeatletter\ifx\SetFigFont\undefined
\def\x#1#2#3#4#5#6#7\relax{\def\x{#1#2#3#4#5#6}}%
\expandafter\x\fmtname xxxxxx\relax \def\y{splain}%
\ifx\x\y   
\gdef\SetFigFont#1#2#3{%
  \ifnum #1<17\tiny\else \ifnum #1<20\small\else
  \ifnum #1<24\normalsize\else \ifnum #1<29\large\else
  \ifnum #1<34\Large\else \ifnum #1<41\LARGE\else
     \huge\fi\fi\fi\fi\fi\fi
  \csname #3\endcsname}%
\else
\gdef\SetFigFont#1#2#3{\begingroup
  \count@#1\relax \ifnum 25<\count@\count@25\fi
  \def\x{\endgroup\@setsize\SetFigFont{#2pt}}%
  \expandafter\x
    \csname \romannumeral\the\count@ pt\expandafter\endcsname
    \csname @\romannumeral\the\count@ pt\endcsname
  \csname #3\endcsname}%
\fi
\fi\endgroup
\mbox{\beginpicture
\setcoordinatesystem units < 1.000cm, 1.000cm>
\unitlength= 1.000cm
\linethickness=1pt
\setplotsymbol ({\makebox(0,0)[l]{\tencirc\symbol{'160}}})
\setshadesymbol ({\thinlinefont .})
\setlinear
%
%
\linethickness= 0.500pt
\setplotsymbol ({\thinlinefont .})
\plot  1.683 26.226  0.445 24.797 /
%
%
\linethickness= 0.500pt
\setplotsymbol ({\thinlinefont .})
\plot  1.683 22.892  0.445 24.320 /
%
%
\linethickness= 0.500pt
\setplotsymbol ({\thinlinefont .})
\plot  1.683 22.892  1.397 24.320 /
%
%
\linethickness= 0.500pt
\setplotsymbol ({\thinlinefont .})
\plot  1.683 22.892  1.968 24.320 /
%
%
\linethickness= 0.500pt
\setplotsymbol ({\thinlinefont .})
\plot  1.683 22.892  2.921 24.320 /
%
%
\linethickness= 0.500pt
\setplotsymbol ({\thinlinefont .})
\plot  1.683 26.226  2.921 24.797 /
%
%
\linethickness= 0.500pt
\setplotsymbol ({\thinlinefont .})
\plot  1.683 26.226  2.064 24.797 /
%
%
\linethickness= 0.500pt
\setplotsymbol ({\thinlinefont .})
\plot  1.683 26.226  1.397 24.797 /
%
%
\put{$\Op P_1$} [lB] at  0.349 24.416
%
%
\put{$\perp$} [lB] at  0.84 24.416
%
%
\put{$\Op P_3$} [lB] at  1.302 24.416
%
%
\put{$\Op P_4$} [lB] at  2.826 24.416
%
%
\put{$\perp$} [lB] at  2.450 24.416
%
%
\put{$\Op P_2$} [lB] at  1.969 24.416
%
%
\put{$\Op 1$} [lB] at  1.62 26.321
%
%
\put{$\Op 0$} [lB] at  1.587 22.511
\linethickness=0pt
\putrectangle corners at  0.349 26.797 and  2.921 22.511
\endpicture}
$$
where:
$$
\Op 0\stackrel{\rm def}{=}[\emptyset]=\left\{\emptyset\right\}\,,\quad
\Op 1\stackrel{\rm def}{=}[M]=\left\{M\right\}\,,\quad
\Op P_j\stackrel{\rm def}{=}[T_j]=\left\{T_j\right\}\,.
$$
\vskip 8mm

\noindent
Of course, in order to model such typical quantum effects like
nonexistence of dispersion free states or violation of Bell's inequality
one has to introduce equivalence classes in a nontrivial way.

\end{document}